# Structure and electronic properties of the $(\sqrt{3}\times\sqrt{3})R30°$ SnAu$_2$/Au(111) surface alloy

M. Maniraj[1,*], D. Jungkenn[1], W. Shi[2,3], S. Emmerich[1,4], L. Lyu[1], J. Kollamana[1], Z. Wei[1,5], B. Yan[2,6], M. Cinchetti[7], S. Mathias[8], B. Stadtmüller[1,4], M. Aeschlimann[1]

[1]Department of Physics and Research Center OPTIMAS, University of Kaiserslautern, 67663 Kaiserslautern, Germany
[2]Max Planck Institute for Chemical Physics of Solids, 01187 Dresden, Germany
[3]School of Physical Science and Technology, ShanghaiTech University, Shanghai 200031, China
[4]Graduate School of Excellence Materials Science in Mainz, Erwin-Schrödinger-Straβe 46, 67663 Kaiserslautern, Germany
[5]College of Materials Science and Engineering, Chongqing University, 400044 Chongqing, PR China
[6]Department of Condensed Matter Physics, Weizmann Institute of Science, Rehovot, 7610001, Israel
[7]Experimentelle Physik VI, Technische Universität Dortmund, 44221 Dortmund, Germany
[8]I. Physikalisches Institut, Georg-August-Universität Göttingen, Friedrich-Hund-Platz 1, 37077 Göttingen, Germany

Abstract

We have investigated the atomic and electronic structure of the $(\sqrt{3}\times\sqrt{3})R30°$ SnAu$_2$/Au(111) surface alloy. Low energy electron diffraction and scanning tunneling microscopy measurements show that the native herringbone reconstruction of bare Au(111) surface remains intact after formation of a long range ordered $(\sqrt{3}\times\sqrt{3})R30°$ SnAu$_2$/Au(111) surface alloy. Angle-resolved photoemission and two-photon photoemission spectroscopy techniques reveal Rashba-type spin-split bands in the occupied valence band with comparable momentum space splitting as observed for the Au(111) surface state, but with a hole-like parabolic dispersion. Our experimental findings are compared with density functional theory (DFT) calculation that fully support our experimental findings. Taking advantage of the good agreement between our DFT calculations and the experimental results, we are able to extract that the occupied Sn-Au hybrid band is of (*s, d*)-orbital character while the unoccupied Sn-Au hybrid bands are of (*p, d*)-orbital character. Hence, we can conclude that the Rashba-type spin splitting of the hole-like Sn-Au hybrid surface state is caused by the significant mixing of Au *d*- to Sn *s*-states in conjunction with the strong atomic spin-orbit coupling of Au, i.e., of the substrate.

# 1. Introduction

The future success to push next generation information technology devices towards the two-dimensional (2D) limit critically depends on our ability to manipulate and control material properties on ever smaller lengthscales, ultimately within a single atomic layer. In this regard, the material class of graphene and its analogs silicene, germanene and stanene are highly promising as they are predicted to be producible in atomically thin films that exhibit exceptional electronic properties.[1,2] However, the possibility to optimize these materials for applications is rather limited as they consist only of one element and preferentially form honeycomb like structure. In contrast, binary systems such as surface alloys consisting of two different elements are much more flexible. Importantly, multiple immiscible elements in their bulk phase can easily form a stable alloy phase upon surface alloying[3] and offer a unique way to tune the geometric and electronic properties of the 2D material system according to the desired functionality[4–12].

For this reason, extensive efforts were devoted to study the electronic properties of 2D surface alloys formed on noble metal surfaces. For fcc(111) surfaces, the combination of heavy metal atoms such as Bi, Pb, Sn, and Sb and noble metal host materials results in the formation of long-range ordered $(\sqrt{3} \times \sqrt{3})R30°$ superstructures in which each third noble metal surface atom is replaced by a heavy metal atom[6,13–27,11]. Most interestingly, the hybridization between both materials results in the formation of a novel 2D surface band structure showing a hole-like Rashba type spin-split state[4,7,28]. The magnitude of the Rashba-type spin splitting thereby depends crucially on the strength of the atomic spin-orbit coupling of the alloy and the host material, on the orbital character of the 2D hybrid band structure[29–32], as well as on the potential landscape of the surface layer[19,33,34]. The latter is determined by the structural properties of the alloyed layer, in particular, by the vertical relaxation of the alloy and host atoms as well as by a potential reconstruction of the entire surface. All these externally controllable parameters offer a highly flexible way to tune 2D materials properties for spintronics applications.[6–9] However, a comprehensive understanding of such surface alloys is still lacking, despite the tremendous amount of experimental and theoretical studies in this field.

In this work, we will focus on a SnAu$_2$ surface alloy formed on a Au(111) single crystal surface. The Au(111) substrate is of particular interest for several reasons. On the one hand, it was recently shown that several elements such as Na, Pb, Ce, La, and Gd can form highly interesting surface alloys on Au(111) with Moiré pattern (*i.e.* complex surface reconstruction) and even with a 2D ferromagnetic phase.[35–41] On the other hand, the intrinsically large spin-orbit coupling (SOC) of Au in comparison to Cu or Ag is expected to lead to an enhanced spin-splitting of the band structure for any alloy formation on the Au surface. In fact, a significantly enhanced Rashba splitting was recently reported for the AuPb/Pb(111) surface alloy.[40] In this context, theoretical studies also revealed the crucial role of a finite *d*-orbital contribution to the primarily *sp*-like surface state of the noble metals for the magnitude of the Rashba-type splitting. The energetic proximity of 5*d*- and 6*s*-orbitals in Au effectively increases the *d*-orbital character in the surface states of Au(111) which is eventually responsible for its significantly larger Rashba spin splitting compared with the surface states of Cu(111) and Ag(111)[29,30,42]. The recently reported absence of any Rashba type-spin splitting of the 2D surface state of a surface alloy formed between Sn and the noble metal Ag perfectly supports this model[18]. To further explore the role of the substrate material for the Rashba spin-splitting of the alloy surface state, we have chosen Sn as the alloy atom. In particular, we aim here to reveal whether the large atomic SOC of Au and the close proximity of d-states and sp-like states in Au(111) are strong enough to mediate the formation of a spin-split surface state for a Sn-Au surface alloy.

In this work, we first report on the lateral order of the SnAu$_2$/Au(111) surface alloy. Using low energy electron diffraction (LEED) and scanning tunneling microscopy (STM), we find the formation of a well-ordered $(\sqrt{3} \times \sqrt{3})R30°$ alloy superstructure which does not lift the herringbone reconstruction of the Au(111) surface. The occupied and unoccupied electronic structure of the SnAu$_2$ surface alloy is investigated by conventional angle-resolved photoemission spectroscopy (ARPES) and angle-resolved two-photon photoemission spectroscopy (AR-2PPE) techniques [27,43]. We find the formation of new surface states in the occupied as well as the unoccupied part of the surface band structure. Most importantly, we reveal that the occupied hybrid surface state shows a hole-like parabolic dispersion with significant Rashba-type spin splitting that is comparable to the splitting of the Au(111) Shockley surface state. The experimental results are discussed in the light of density functional theory (DFT) calculations which allow us to reveal the origin of the spin-splitting of surface bands of the Sn-Au surface alloy.

## 2. Methods

All sample preparation steps and experiments were performed under ultrahigh vacuum conditions. The Au(111) surface was prepared by repeated sputtering and annealing cycles. The quality of the clean surface was confirmed by the existence of sharp diffraction spots that correspond to the well-known Au-herringbone reconstruction in LEED (see Fig. 1(a)), and by the appearance of the Rashba splitting of the surface state in ARPES.[44] Subsequently Sn (purity 99.9999%) was deposited onto the clean surface at an elevated sample temperature of about 650 K using a water cooled evaporator in which a PBN crucible containing Sn was heated to nearly 1100 K.[45,46] During the evaporation, the pressure remained better than $5\times10^{-10}$ mbar. The success of the sample preparation was confirmed by LEED. All photoemission, LEED and STM experiments were performed in separate chambers at room temperature with base pressures better than $2\times10^{-10}$ mbar. Conventional ARPES experiments were performed using a Specs 150 hemispherical analyzer using He I light sources and the fourth harmonic of a Ti:Sa laser (5.9 eV). The unoccupied electronic structure was investigated with AR-2PPE with the second harmonic of the Ti:Sa laser at energies of 2x 3.1 and 2x 2.95 eV. The p-polarized beams from the laser were used at an incident angle of 45° with respect to the analyzer's optical axis. The total energy and angular resolution was determined to be better than 80 meV (as a result of combination of light source, analyzer settings, and room temperature measurements) and 0.3°, respectively. All STM measurements were performed using a chemically etched and *in-situ* sputter-annealed tungsten tip.[47,48] The STM images were recorded in constant current mode and subsequently processed using WSxM software.[49]

The density functional theory calculations were performed using the Vienna *Ab-initio* Simulation Package (VASP)[50]. The interactions between the valence electrons and ion cores were described by the projector augmented wave method[51,52]. The electron exchange and correlation energy was treated by the generalized gradient approximation with the Perdew-Burke-Ernzerhof functional[53]. The kinetic energy cutoff of the plane-wave basis was set to 248.3 eV as default. The first Brillouin zone sample used the Γ-centered 6×6×1 *k* points. The structures were optimized until the forces on the atoms were less than 5 meV Å$^{-1}$. The Au (111) surface was described by a periodic slab separated by 20 Å vacuum. 31 layers of Au atoms were included with the center 25 layers fixed as the bulk crystal structure while the top and bottom 3 layers were relaxed. One monolayer of SnAu$_2$ atoms was adsorbed on each top and bottom side of the Au slab. The herringbone reconstruction of the of the Au(111) could not be considered in our calculations due to its large unit cell and the correspondingly extremely large number of atoms per unit cell. All the bands are rigidly shifted down by 100 meV to match the ARPES data.

## 3. Experimental Results

We start our discussion with the atomic structure of the SnAu$_2$/Au(111) surface alloy. The LEED pattern of the clean sample surface is shown in Fig. 1(a). It reveals the typical LEED pattern of the Au(111) surface with six sharp diffraction spots arranged in a hexagonal pattern. Additionally, the specular reflection is surrounded by the six diffraction spots of the herringbone reconstruction, which are clearly visible in the inset of Fig. 1(a). After deposition of approximately one third of a monolayer of Sn on the Au(111) surface at a sample temperature of 650 K, the LEED pattern changes and we observe the characteristic diffraction pattern of a $(\sqrt{3} \times \sqrt{3})R30°$ superstructure (hereafter referred to as R3 phase), which is depicted in Fig. 1(b). This diffraction pattern is observed for (almost) all 2D surface alloys formed on fcc(111) noble metal surfaces[14,15,18,40] and hence provides the first evidence that Sn indeed forms a surface alloy on Au(111). Most interestingly, the diffraction pattern of the R3 phase still reveals the characteristic diffraction spots of the herringbone reconstruction at identical positions in momentum space suggesting that the surface reconstruction is not lifted and its periodicity is not altered by the adsorption of Sn atoms, see inset in Fig. 1(b). Since the herringbone reconstruction of the Au(111) surface is the result of an uniaxial compression of surface atoms along the <110> direction, we can conclude that the exchange of Sn and Au atoms in the first layer does not reduce the surface stress along the <110> direction. To investigate the uniformity and local atomic structure of the alloy surface we turn to the local probe technique STM. In the large area constant-current STM image in Fig. 1(c), we find smooth and island-free terraces of the well-ordered film of the R3 phase which confirms the formation of a high-quality and uniform surface film with low defect density. In addition, the characteristic double-stripe-like pattern of the herringbone reconstruction is also clearly visible in Fig. 1(c) (see dashed-double-lines)[54–57]. The distance between two adjacent double-stripe lines, i.e., the separation between two subsequent hcp regions, is 7.5 nm and hence identical to the one of the bare Au(111) surface. Thus, both LEED and STM experiments confirm that Sn adsorption does not lift the herringbone reconstruction and its periodicity remains unchanged. The atomic structure can be resolved in the detailed high-resolution STM image that is shown in Fig. 1(e). It is dominated by bright protrusions that are arranged in a hexagonal lattice with a periodicity of 5 Å. This distance corresponds to the lattice constant of a R3 phase, i.e., the $\sqrt{3}a$ superlattice of the SnAu$_2$/Au(111) surface ($a$ = 2.88 Å) and hence is fully in line with our LEED results. A structural model of the R3 phase based on LEED and STM (without herringbone reconstruction) is illustrated in Fig. 1(f) and 1(g). Our DFT data only reveal a marginal vertical relaxation (z= 0.06 Å) of the Au and Sn surface atoms with respect to the truncated lattice plane of the Au(111) surface. This is severely different compared to (almost) all surface alloys formed on the two noble metals Cu(111) and Ag(111) for which the significantly different atomic radius of the alloy and host atoms leads to a vertical relaxation of the alloy atoms with respect to the surface plane of the host material[34].

The second obvious difference between our findings and those of the surface alloys formed on Cu(111) and Ag(111) is the existence of the herringbone surface reconstruction. Similar surface reconstructions or Moiré pattern are commonly reported for various elements (Na, Pb, Ce, La, and Gd) forming a R3 phase on an Au(111) substrate[35–39,41]. This indicates that the surface stress of the reconstructed Au(111) surface is not substantially modified by the adsorption of alloy atoms.[58] In particular, the persistence of the herringbone reconstruction upon the adsorption of Sn will have important consequences for the properties of the surface alloy. On the one hand, the surface reconstruction can result in an additional in-plane surface potential gradient which would strongly enhance any Rashba-type spin-splitting of the Sn-Au surface alloy band structure. On the other hand, the high reactivity of the elbows and the point dislocations of the herringbone reconstruction can act as nucleation centers for the growth of organic and inorganic adsorbates on this functionalized surfaces[59–62] or can influence the efficiency of surface-induced chemical reactions.[59,63–67]

We now turn to the electronic structure investigated by photoemission spectroscopy. Upon the adsorption of Sn on Au(111), the Rashba-type spin-split surface state of Au(111)[44] is totally suppressed and replaced by a set of new hole-like parabolically dispersing bands around the $\bar{\Gamma}$ point that are marked as S1 and S2 in Fig. 2(a) and 2(d). The band structure of the R3 phase around the $\bar{\Gamma}$-point is similar for both high symmetry directions ($\overline{\Gamma K'M}$ and $\overline{\Gamma M'K}$) of the Au(111) surface Brillouin zone (see inset in Fig. 2(b) for the Brillouin zone scheme). For a quantitative analysis of the band dispersion of S1 and S2, we repeated the ARPES measurement with monochromatic He $I_\alpha$ light. The corresponding data are shown in Fig. 2(b) and 2(e) for both high symmetry directions. For better visibility of the band dispersion, the ARPES data are shown as a waterfall plot in Fig. 2(c) and 2(f). Our high resolution ARPES data reveal that both bands S1 and S2 consist of two subbands that are symmetrically shifted in momentum space around the $\bar{\Gamma}$-point at about ±0.2 Å$^{-1}$ and ±0.4 Å$^{-1}$ at the Fermi level, respectively. Such a momentum space shift of the newly formed alloy surface state has been observed for various surface alloys such as Pb, Bi, and Sb on Ag(111) and Cu(111) [6,7,12–24,33,34,68] and is characteristic for a Rashba-type spin splitting of the alloy surface state[4,7,28,44]. However, our results are qualitatively different compared to the Sn-Ag surface alloys, for which no Rashba-type spin-splitting was observed[18]. The results of our quantitative analysis of the dispersions of S1 and S2 is shown in Fig. 2h. Fitting allows us to estimate the Rashba parameter for both bands using the relation $\alpha_R = \hbar^2 k_0/m^*$, and $E_R = \hbar^2 k_0^2/2m^*$, where $\hbar$ is the Planck's constant, m* the effective mass of the electron, and $k_0$ the momentum offset of spin split parabolas as shown in inset of Fig.2(h). We obtain a Rashba-parameter $\alpha_R$= 0.4±0.002 eVÅ$^{-1}$ (Rashba-constant) and $E_R$= 4±2 meV (with effective mass $m^*/m_e$ =0.32±0.01 and momentum offset $\Delta k_\parallel$= 0.02±0.02 Å$^{-1}$) for the band S1 and $\alpha_R$= 0.44±0.001 eVÅ$^{-1}$ (Rashba-constant) and $E_R$= 6±2 meV (with effective mass $m^*/m_e$ =0.43±0.01 and momentum offset $\Delta k_\parallel$= 0.025±0.02 Å$^{-1}$) for S2. Here, we would like to point out that our parabolic fitting performed for S2 in Fig. 2(h) shows the band crossing at about 0.9 eV above the Fermi energy. However, as will discussed in the following, our DFT results reveal the band crossing for S2 around 2.5 eV above $E_F$. Therefore, we will focus in our quantitative discussion on the experimental Rashba parameter extracted for S2. All values obtained for S1 and S2 are comparable to the Rashba-parameters of the bare Au(111) surface state ($m^*$ = 0.25 $m_e$ and $\Delta k_\parallel$= 0.023 Å$^{-1}$)[44,69,70].

In the second part of our ARPES study, we address the occupied and unoccupied band structure of the R3 phase using different laser light sources. Figure 3(a) shows ARPES data of the occupied band structure obtained with hν= 5.9 eV. The spectrum is dominated by two features that are labeled as S1 and M. The band S1 appears symmetrically around the $\bar{\Gamma}$-point at about ±0.2 Å$^{-1}$ at the Fermi level and follows the same dispersion as the band S1 observed with He I radiation. We hence attribute it to the same band. The lack of $k_z$-dispersion of this state confirms the identification of this state as a surface state. The other distinctive feature is the extremely bright parabolic-like band labeled as M. This band is assigned to the Mahan cone formed by a resonant *sp-sp* bulk transitions for the Au(111) substrate.[71]

To investigate the unoccupied band structure, we have performed monochromatic AR-2PPE measurements using a photon energy of hν$_1$=2.95 eV. The corresponding AR-2PPE data are shown in Fig. 3(b). In addition to the band S1 observed in conventional ARPES, we find two additional sets of bands around the $\bar{\Gamma}$-point that are labeled as S3 and S4. Note that S1 still contributes to the AR-2PPE data due to a strongly detuned transition from the occupied initial state S1.

In order to confirm that both features S3 and S4 are signatures of the unoccupied band structure and not initial or final state effects of the 2PPE process, we repeated the AR-2PPE measurement with another photon energy of hν$_2$= 3.1 eV. A closer look at the AR-2PPE data reveals that the kinetic energy of S1 shifts by 2×|hν$_1$–hν$_2$| as expected for an occupied band. In contrast, the kinetic energy of S3 and S4 increases by 1×|hν$_1$–hν$_2$| indicating that S3 and S4 are bands of the unoccupied part of the band structure. However, the origin of both bands is still not unambiguously clear. At first glance, it seems likely that S3 and S4 are two Rashba-type spin-split parabolic bands that are shifted by a constant value in momentum space. This question will be addressed in the following by DFT calculations.

## 4. Band structure calculations based on DFT

To gain insight into the origin of the Rashba-type spin splitting of the band S1, we now turn to our density functional theory band structure calculations. We start our discussion with the surface band structure projected onto the topmost SnAu$_2$ surface alloy layer along the $\overline{\Gamma M' K}$ direction of the surface Brillouin zone shown in Fig. 4. We can clearly identify all four experimentally observed hole-like bands S1, S2, S3, and S4 in our calculations even without considering SOC (Fig. 4(a)). S1 appears as a single band centered at the $\bar{\Gamma}$-point of the surface

Brillouin zone and shows a hole-like parabolic dispersion in agreement with our experiment. In the unoccupied part of the band structure, we find two bands with almost linear dispersion. These bands qualitatively resemble the dispersion of both features S3 and S4 observed in our AR-2PPE experiment (Fig. 3(b) and 3(c)). The finite energetic difference of S3 in DFT and the AR-2PPE experiment (approx. 0.5 eV) can be attributed to the unaccounted self-energy correction in the DFT calculations for unoccupied states, which is frequently observed when comparing experiment and DFT[72,73]. Despite this quantitative difference, the mere existence of two separated bands S3 and S4 in our DFT calculation without SOC allows us to conclude that S3 and S4 are not two branches of the same Rashba-type spin split state, but rather two individual bands.

When SOC is included in the DFT calculations, we find only marginal, but characteristic changes in the surface band structure of the $SnAu_2$ surface alloy as shown in Fig. 4(b). S1 splits into two parabolic sub-bands which are symmetrically displaced from the $\overline{\Gamma}$-point by about 0.2 Å$^{-1}$ at the Fermi level. The calculated band structure perfectly describes the experimentally observed band dispersion of S1 with its momentum offset of about 0.2 Å$^{-1}$. It also clearly reveals the characteristic fingerprint of a Rashba-type split surface state which becomes even more evident when considering the spin-character of the branches of S1 in the spin-resolved band structure in Fig. 4(c) and its inset. The latter was extracted from our DFT calculation with SOC and the red and blue dots indicate the spin-character of the corresponding bands projected onto two opposite spin orientations in the surface plane. We find that both branches of S1 possess opposite spin directions as expected for Rashba type splitting[4,7,74].

In contrast, we do not observe a significant splitting of S2, S3 and S4 along the $\overline{\Gamma M'K}$ direction, nor a clear spin-polarization of these bands when including SOC in our calculation. For the unoccupied bands S3 and S4, this finding is completely in line with our AR-2PPE data and hence again confirms that S3 and S4 are two distinct bands and not two (spin-polarized) branches of the same band. While our calculation is able to qualitatively describe the dispersion of S2, it fails to predict the splitting of S2 in two parabolic sub-bands along $\overline{\Gamma M'K}$ direction observed in our ARPES experiment, Fig. 2(b) and (c).

This discrepancy between experimental data and theoretical predictions for S2 can be resolved when considering the result of our band structure calculation along the $\overline{\Gamma K'M}$ direction. The corresponding spin-integrated and spin-resolved surface projected band structures are shown in Fig. 4(d) and (e). Most interestingly, the spin-splitting of the surface alloy hybrid bands S1 and S2 along the $\overline{\Gamma K'M}$ direction is exactly opposite compared to the $\overline{\Gamma M'K}$ direction discussed above. While S1 no longer reveals any spin splitting along the $\overline{\Gamma K'M}$ direction, the spin degeneracy of S2 is lifted along this high symmetry direction. Our band structure calculations shows a clear sub-band splitting for S2 with a maximum value of $\Delta k_\parallel \approx$ 0.03 Å$^{-1}$ at a binding energy of -1.2 eV, see inset in Fig. 4(e). This different spin splitting of the hybrid bands S1 and S2 along both high symmetry directions can be attributed to a hexagonal warping of the Fermi contour which has already been observed for spin-textured surfaces states in topological materials and Rashba-type surface alloys[13,32,75–81]. In our adsorbate system, the Fermi surface warping can be explained by the threefold (p3m1) symmetry of the SnAu$_2$/Au(111) surface alloy with an additional in-plane inversion asymmetry[13,79] leading to the Fermi contour of S1 and S2 as schematically shown in Fig. 4(f).

Although our band structure calculation reveals a characteristic spin-splitting of S1 and S2 in different high symmetry directions in momentum space, it cannot explain the isotropic momentum space splitting of S1 and S2 observed in our ARPES experiment. This final deviation between experiment and theory is attributed to the herringbone reconstruction of the surface layer of the SnAu$_2$ surface alloy which is not considered in our DFT calculation. The vertical corrugation of the (22×√3) herringbone reconstruction results in a uniaxial in-plane anisotropy of the surface which lowers the symmetry of the surface from a six-fold symmetry of the SnAu$_2$ superlattice to a two-fold symmetry. This reduction of the alloy surface symmetry together with p3m1 symmetry of the Au(111) substrate leads to an azimuthal averaging of the Fermi surface and hence to an identical band structure of S1 and S2 for both high symmetry directions of the surface Brillouin zone.

## 5. Discussion

The generally good agreement between experiment and theoretical calculations further allows us to identify the individual orbital character of the hybrid bands S1, S2, S3 and S4 from DFT. The corresponding orbital-projected band structure of the *d*, *p*, and *s* orbitals of Au and Sn $\overline{\Gamma M'K}$ and $\overline{\Gamma K'M}$ high symmetry direction are shown in Fig. 5(a)-(e). The spatial coordinates x, y, and z of the orbitals correspond to the directions of a right-handed Cartesian coordinate system in which the x- and y-axis are oriented along the $[\bar{1}10]$ direction (corresponding to the $\overline{\Gamma M'K}$ direction in momentum space) and the $[\bar{1}\bar{1}2]$ direction (corresponding to the $\overline{\Gamma K'M}$ direction in momentum space) as shown in Fig. 2(g), respectively. The positive z-axis marking the direction of the surface normal towards the vacuum. We find that the spectral features of all orbitals oriented in the surface plane (x-, y-direction) are the dominant orbitals of S1-S4 and their relative contributions are almost identical.

For both $\overline{\Gamma M'K}$ and $\overline{\Gamma K'M}$ high symmetry direction (Fig. 5(a)-(e)), the band S1 consists mainly of Sn $s$-like, and Au $d_{xy}$-, and $d_{x^2}$-like orbitals and is hence the result of a direct hybridization between the Sn and the Au bands. The relative individual orbital contributions can be extracted from the weight-projected band structure. The following values for S1 were extracted from the band structure projection along the $\overline{\Gamma M'K}$ direction which reveals the experimentally expected spin splitting of S1. Around the $\bar{\Gamma}$-point ($k_\parallel = \pm 0.4$ Å$^{-1}$), we find $s$-orbital contributions between 30-50% as well as $d_{xy}$-, $d_{x^2}$- and $p_y$-orbital contributions of about 5–15 %, 3–5 % and 3–6 %, respectively, see Fig. 5(f, top). Along these lines, we only observed a marginal contribution of $p_z$ and $d_{z^2}$-orbitals to the formed hybrid S1 surface states. Interestingly, the relative contributions of all orbitals to S1 are not constant, but vary with the crystal momentum ($k_\parallel$) as shown in Fig. 5(f, top). The contribution of the $s$-orbital is almost constant around the $\bar{\Gamma}$-point before decreasing monotonically for $k_\parallel$-values larger than 0.2 Å$^{-1}$. In contrast, the contributions of both $d$-orbitals increase with increasing $k_\parallel$. This increase coincides with an enhanced energy splitting $\Delta E(k_\parallel)$ between bands of opposite spin character providing clear evidence that the spin splitting of the band S1 crucially depends on the contribution of Au $d$-orbitals to the $s$-like states of Sn. Interestingly, these orbital contributions of $s$-, $p_x$-, $p_y$-, $d_{xy}$-, and $d_{x^2}$- states decrease along the $\overline{\Gamma K'M}$ direction in the range of $|k_\parallel| = 0.2$–$0.5$ Å$^{-1}$. The projected weights of the $d$-orbitals along the $\overline{\Gamma K'M}$ direction are 50 % smaller compared to the corresponding values along $\overline{\Gamma M'K}$ direction [≈10 % (Fig. 5(f, top)) vs. ≈20 % (Fig. 5(f, bottom))]. This significantly smaller $d$-band contribution to S1 is hence responsible for the neglectable spin-splitting of S1 along the $\overline{\Gamma K'M}$ direction.

Similarly strong contributions of $d$-like orbitals to the Rashbas-type surface states have also been reported for the spin-split surface state of the bare Au(111) surface and have been identified as one of the main ingredients for its spin-splitting [29,30,42,82]. Theoretically, it was even shown that the surface state of Au(111) consists of *sp*-like bands (50 % of *s*- and 80 % of *p_z*-orbitals) with significant contributions of $d_z^2$, $d_{xz}$, $d_{yz}$ orbitals (in total 30–50 %)[29,30,42,82]. In analogy, we can also conclude that the spin-splitting of the Sn-Au hybrid band S1 is mainly caused by the strong atomic spin-orbit coupling of Au as well as by the mixing of *d*-orbitals of Au with *s*-orbitals of the alloy atom Sn.

Interestingly, the existence and absence of the spin-splitting of S2 is completely opposite to the one of S1 for both high symmetry directions $\overline{\Gamma M'K}$ and $\overline{\Gamma K'M}$, i.e., S2 is spin-degenerated along the $\overline{\Gamma M'K}$ direction, while it reveals a finite spin splitting along the $\overline{\Gamma K'M}$ direction. In analogy to our discussion above, this different behavior is directly correlated to the orbital character of S2 along both high symmetry directions. In particular, S2 exhibits mainly $d_{xy}$-, and $d_x^2$-orbital character along the $\overline{\Gamma M'K}$ direction (see Fig. 5(f, top)) with only a neglectable contribution from *s*-orbitals preventing any spin splitting of S2. In contrast, the hybridization of *d*- and *s*-orbitals along the $\overline{\Gamma K'M}$ direction leads to a clear spin splitting for S2. Thus we conclude that the hexagonal warping of the Fermi contour for both S1 and S2 must be and intrinsic band structure effect of the SnAu$_2$ surface alloy and is not induced by the existence of the herringbone reconstruction.

Turning to the unoccupied part of the band structure, we find that S3 and S4 originate predominantly from *p*- states of Sn and *d* states of Au, Fig. 4(d)–(g). The different character of these Sn+Au *d*-, *p*-like hybrid states in unoccupied band structure compared to *s*-, *d*-like hybridization of occupied states could also be the origin of the negligible spin-splitting of S3 and S4 observed both in AR-2PPE as well as in our DFT calculations. In fact, it has been reported for the case of Au(111) surface state that the *p*-orbital can have a negative effect on the Rashba splitting. The p-orbital contribution can counterbalance the effect of the *d*-band contributions which results in a negligible splitting of the surface state[29].

In the light of previous investigations of surface alloys, our findings for the SnAu$_2$ surface alloy appear rather exceptional. We observe a significant spin-splitting of the hybrid surface state (S1) of (*s*, *d*)-orbital character despite the weak atomic SOC of the alloy atoms and the missing structural relaxation of the surface atoms Sn and Au. Both aspects are clearly different for other heavy metal-noble metal surface alloys such as Bi and Pb on Ag(111) and Cu(111)[13,19,21] for which the surface state is derived from *s*- and *p*-like orbitals and the Rashba-type spin-splitting is mediated by the large SOC of the heavy metal atoms and their vertical structural relaxation with respect to the surface plane.

In the wide family of surface alloys, we find the most similarities between our material system SnAu$_2$ and the SnAg$_2$ surface alloy.[18] The latter Sn-Ag adsorbate system reveals a similar hybrid surface state with almost identical orbital character, i.e., Sn *s*-like and Ag *d*-like orbitals. However, the most striking difference between both surface alloys is the negligible spin-splitting of the Sn-Ag hybrid surface state observed in the center of the surface Brillouin zone. This difference is attributed to the different strength of the atomic SOC of Ag and Au in both surface alloys and the close energetic proximity of *d*- and *s*-orbital in Au substrate, but not to the absence of a herringbone reconstruction for the SnAg$_2$ surface alloy. The latter conclusion is based on the apparently neglectable role of the herringbone reconstruction and subsequent in-plane potential gradient for the Rashba-type spin-splitting of the Sn-Au surface state S1. This is based on the almost perfect quantitative agreement between our ARPES data and our band structure calculations which were performed for an unreconstructed surface alloy (i.e., without herringbone reconstruction). This finding is also in line with the neglectable influence of the herringbone reconstruction for the magnitude of the Rashba spin splitting of the Au(111) Shockley surface state[44,83].

Along these lines, we can conclude that the Rashba-type spin-splitting of the Sn-Au surface state is the result of a (*s*, *d*) like-hybridization and the strong SOC of the Au atoms. This finding is highly interesting since *d*-band induced Rashba type splitting has not been observed so far for surface alloys even though it is known for other material systems such as bare and adatoms covered metal surfaces[16,84–88]. In this aspect, the Sn-Au adsorbate system investigated here opens a new avenue to extend our knowledge from (*s, p*) to (*s, d*) band-mediated Rashba-split states.

## 6. Conclusion

In this work, we have investigated the geometric and electronic structure of an ordered SnAu$_2$ surface alloy on the Au(111) surface using a variety of experimental and theoretical techniques. The combination of LEED and STM reveals the formation of a smooth and long-range ordered SnAu$_2$ surface alloy with a $(\sqrt{3} \times \sqrt{3})R30°$ superstructure, in agreement with similar $A_1B_2$/B(111) surface alloys on noble metal surfaces. Most interestingly, the herringbone reconstruction of the bare Au(111) surface is not lifted by the adsorption of Sn. Turning to the electronic properties, we find that the occupied part of the electronic band structure of the $SnAu_2$ surface alloy is dominated by a new hybrid surface state showing a hole-like parabolic dispersion and a symmetric momentum space splitting around the $\bar{\Gamma}$-point. The latter is consistent with a Rashba-type spin splitting of this new hybrid state. Moreover, the unoccupied band structure reveals two additional Sn-induced bands with almost linear dispersion, but without any indication of a Rashba type spin-splitting. Employing DFT calculations, we were able to show that all new hole-like states of the $SnAu_2$ surface alloy arise due to hybridization between *s*- and *p*-like states of Sn and Au *d*-bands. The spin splitting of this new hybrid surface state is attributed to the strong atomic SOC of Au and to the mixing of Au *d*-orbitals with Sn *s*-like orbitals. In contrast, no spin splitting was observed for the unoccupied hybrid bands consisting of Sn *p*-like and Au *d*-like orbitals neither in experiment nor in DFT calculations. Considering that no spin-split surface states were observed for similar Sn-Ag surface alloys, our comprehensive investigation demonstrates that a Rashba-type spin splitting of hybrid surface states of binary surface alloys can also be induced solely by the strong SOC of the substrate material (in our case Au) and not only by the strong SOC of heavy metal adsorbates.

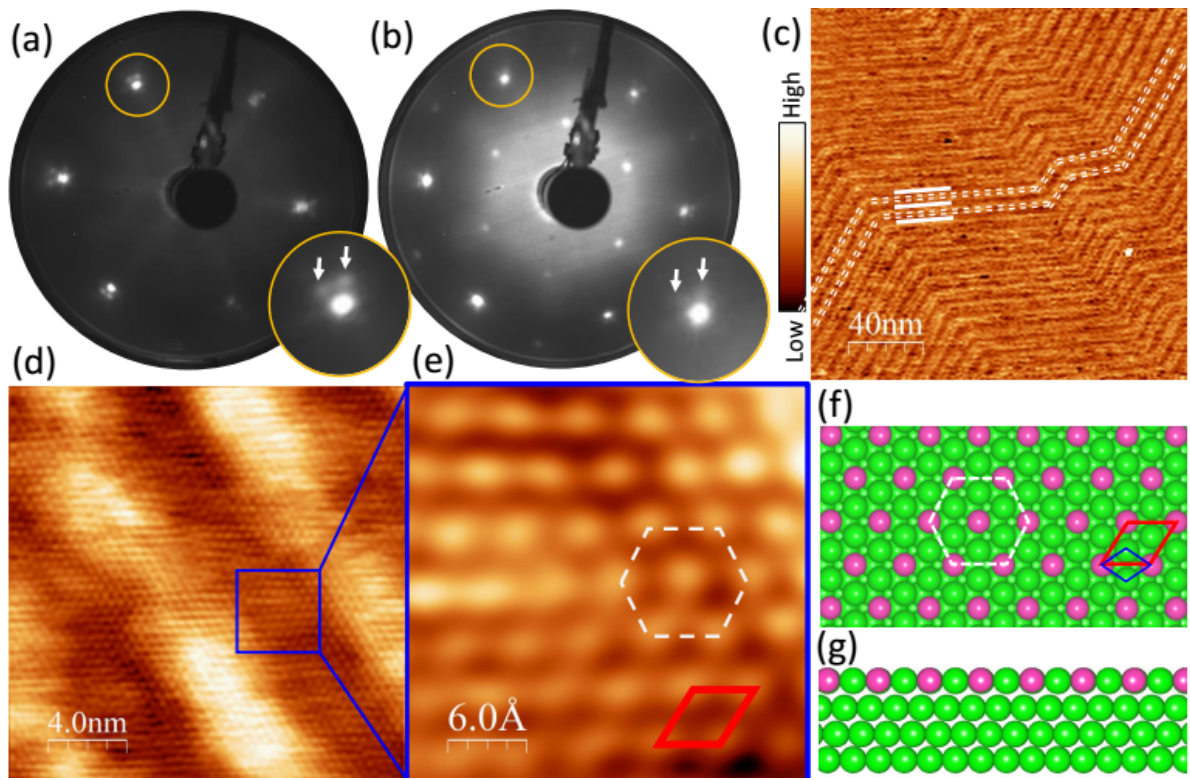

**Figure 1:** (color online)

LEED pattern of (a) bare Au(111) and (b) the R3 phase of Sn/Au(111). Insets in (a) and (b) show the expanded view of the region marked by a yellow colored circle in the main figures. Two of the herringbone reconstruction induced superstructure spots are marked by arrows in the expanded view. (c) Large area (200×200 nm$^2$) constant-current STM image of the R3 phase measured with U= 500 mV, I= 0.1 nA. (d) High resolution (20×20 nm$^2$) STM image measured with U= 80 mV, I= 0.5 nA, and (e) selected area STM image from square marked in (d). The stripe-like pattern of the herringbone reconstruction is marked by dashed-double-lines in (c). Thick continuous-lines in (c) indicate the *fcc* stacking position. The *hcp* region is marked by double-dashed-lines. The distance between the *fcc* stacking regions (thick continuous-lines) is nearly 7.5 nm. All STM images are acquired at 300 K. (f) Top view and (g) side view of a simple hard-sphere ball model of the R3 phase Sn/Au(111) (the herringbone reconstruction is not considered in the model). Pink and green colored spheres correspond to Sn and Au atoms, respectively. The unit cell of bare Au(111) and the R3 phase of Sn/Au(111) are indicated by blue and red colored lines, respectively.

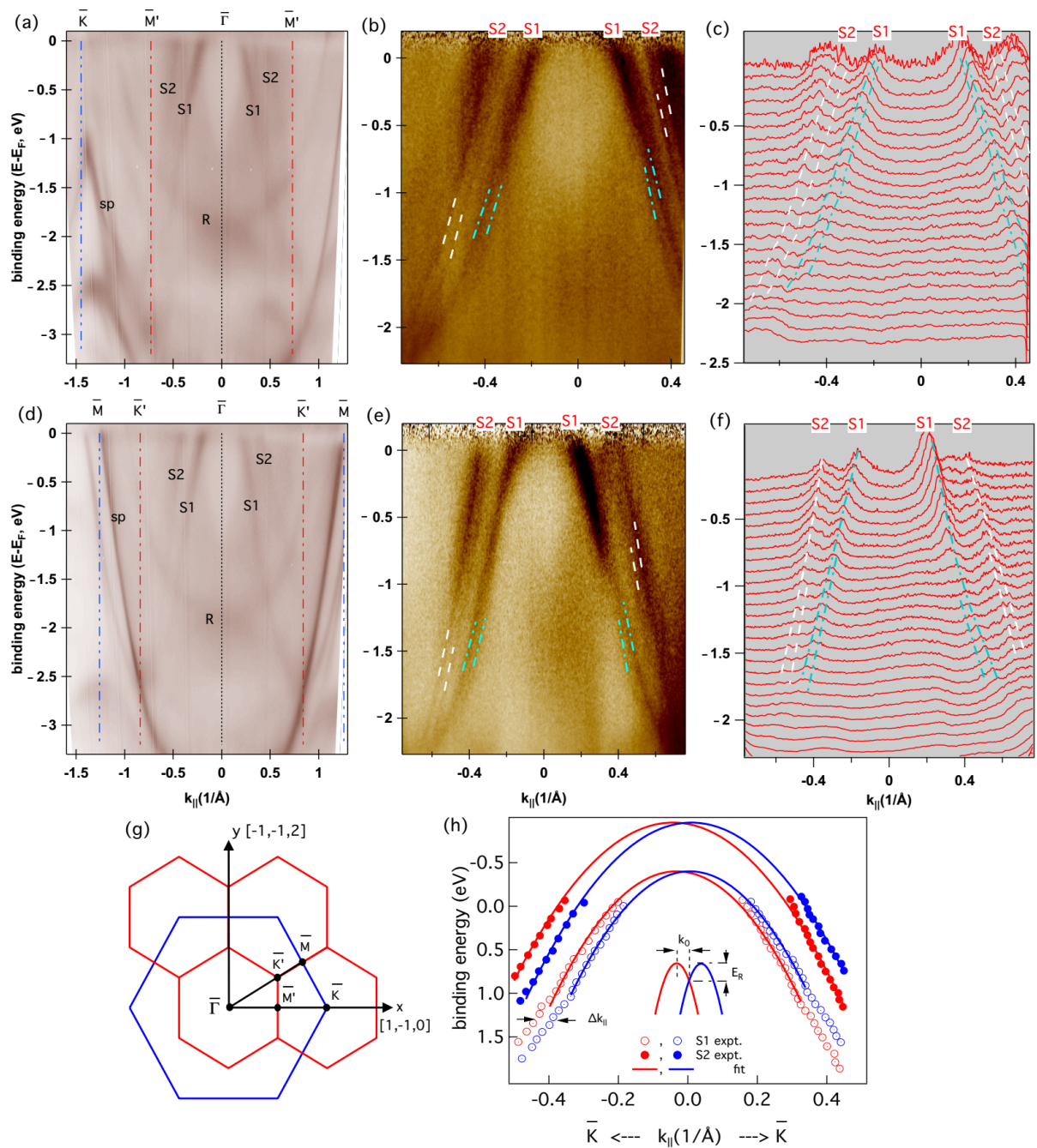

**Figure 2:** (color online)

Angle-resolved photoemission map of the R3 phase of Sn/Au(111) around the $\overline{\Gamma}$-point: (a)–(c) measured along $\overline{\Gamma M'K}$ and (d)–(f) along $\overline{\Gamma K'M}$ direct contemporary ions. Both (a) and (d) are recorded using non-monochromatized He I light. The weaker features "R" in (a) and (d) denote replicas of the bulk band structure due to photoexcitation by the He Iß emission line of the non-monochromatized He I light source. (b) and (e) are recorded using monochromatized He I$_\alpha$ light. (c) and (f) are waterfall plots of images shown in (b) and (e), respectively. The lines in (b), (c) and (e), (f) are a guide to the eye. The ARPES images are obtained by normalizing each angular channel by the averaged energy distribution curve of the corresponding ARPES image. (g) The surface Brillouin zone of the bare Au(111) surface and the R3 phase of Sn/Au(111) are shown in blue and red hexagons, respectively. The x, y axis, and corresponding crystallographic directions in real and momentum space are included for clarity. The positive z axis is pointing normal to the sample surface and is directed towards the vacuum. Panel (h) shows experimental data points of the dispersion of S1 and S2 extracted from (b). The parabolic fitting model of the band dispersion of all branches of S1 and S2 are also included as red and blue solid curves. The inset of (h) shows the definition of $E_R$, $\Delta k_\parallel$, and $k_0$ used for the estimating the Rashba parameter.

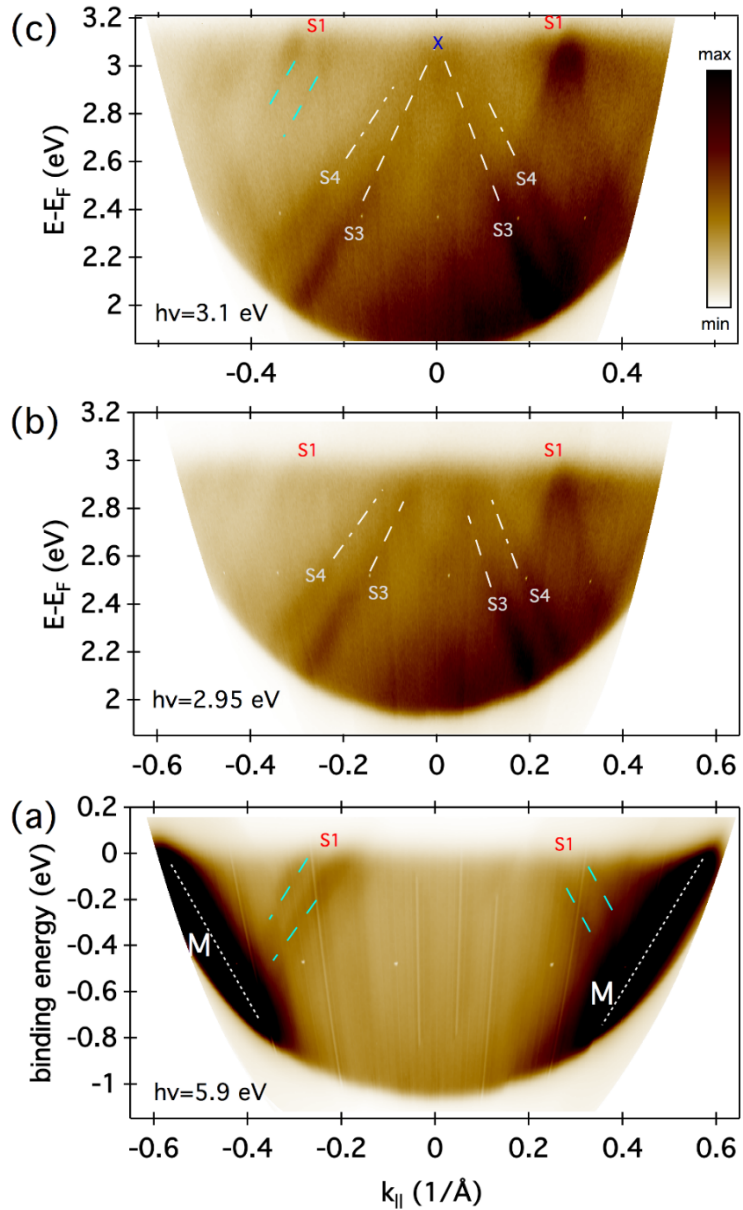

**Figure 3:** (color online)

Experimentally measured occupied and unoccupied electronic states of the R3 phase of the Sn/Au(111) surface along the $\overline{\Gamma M'K}$ direction, respectively. The data in panel (a) were recorded using hν=5.9 eV, (b) and (c) using 2x 2.95 and 2x 3.1 eV photon energy, respectively. All lines in the figure are a guide to the eye. The intensity range for each individual image was selected such that as many features as possible are readily visible.

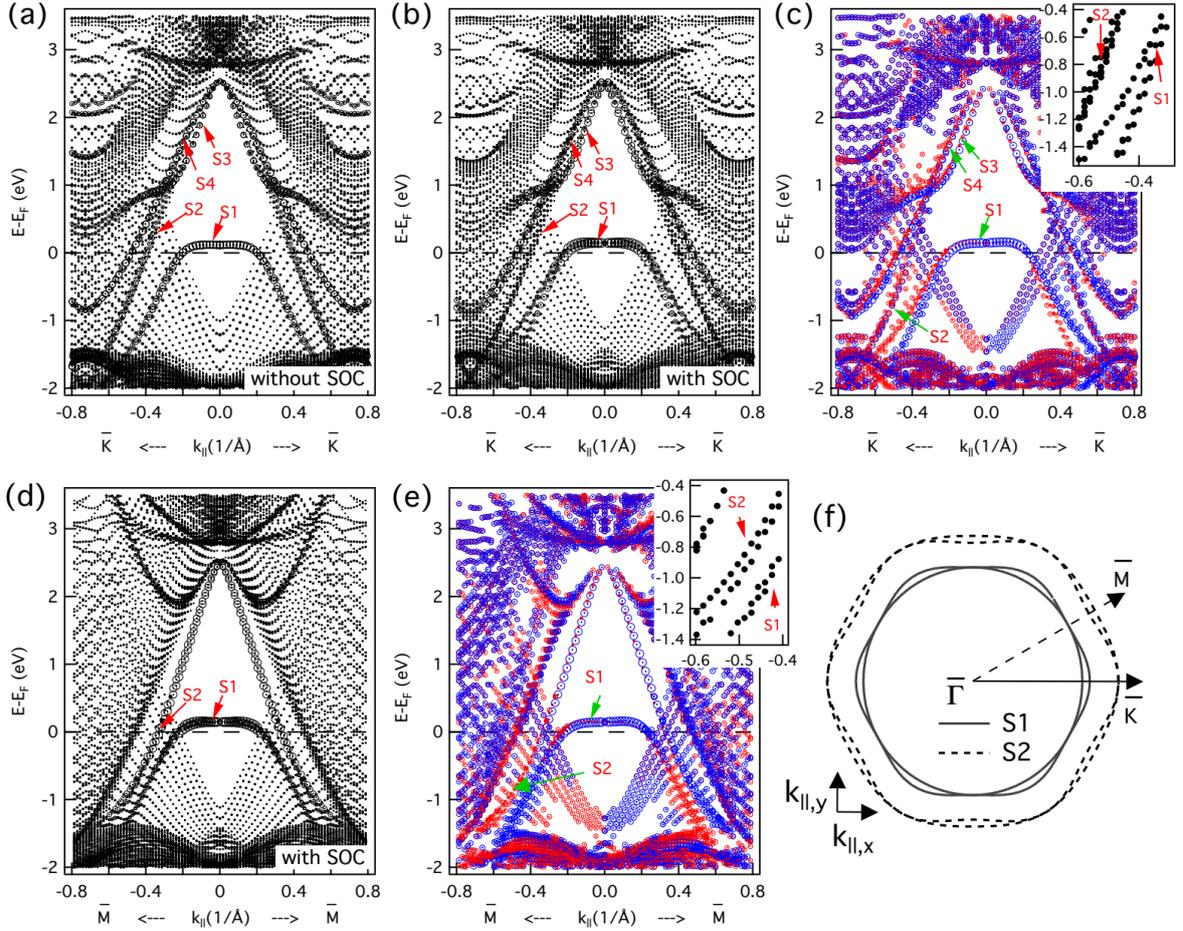

**Figure 4:** (color online)

The surface-, orbital- and spin-projected band structure of the R3 phase of SnAu$_2$/Au(111). Panels (a)-(c) show the band structure along $\overline{\Gamma M'K}$ and (d)-(e) along the $\overline{\Gamma K'M}$ high symmetry direction. All these projected band structures contain only contributions originating from the topmost SnAu$_2$ alloy layer. Panel (a) displays the band structure calculated without spin-orbit coupling (SOC). Panels (b), (c) and (d), (e) depict the spin-integrated and spin-projected band structure with SOC, respectively. In the spin-polarized band structures, the red and blue colors indicate two opposite spin directions in the surface plane. The insets in (c) and (e) show an enlarged view onto the sub-band splitting of the bands S1 and S2. The spin character is not considered in the insets for clarity. (f) Simplest schematic of the hexagonal warping of the bands S1 and S2 at a larger binding energy of nearly -1.2 eV where the splitting is visible for both S1 and S2 in our band structure calculations shown in (b)-(e) and Fig. 5 (see text for details).

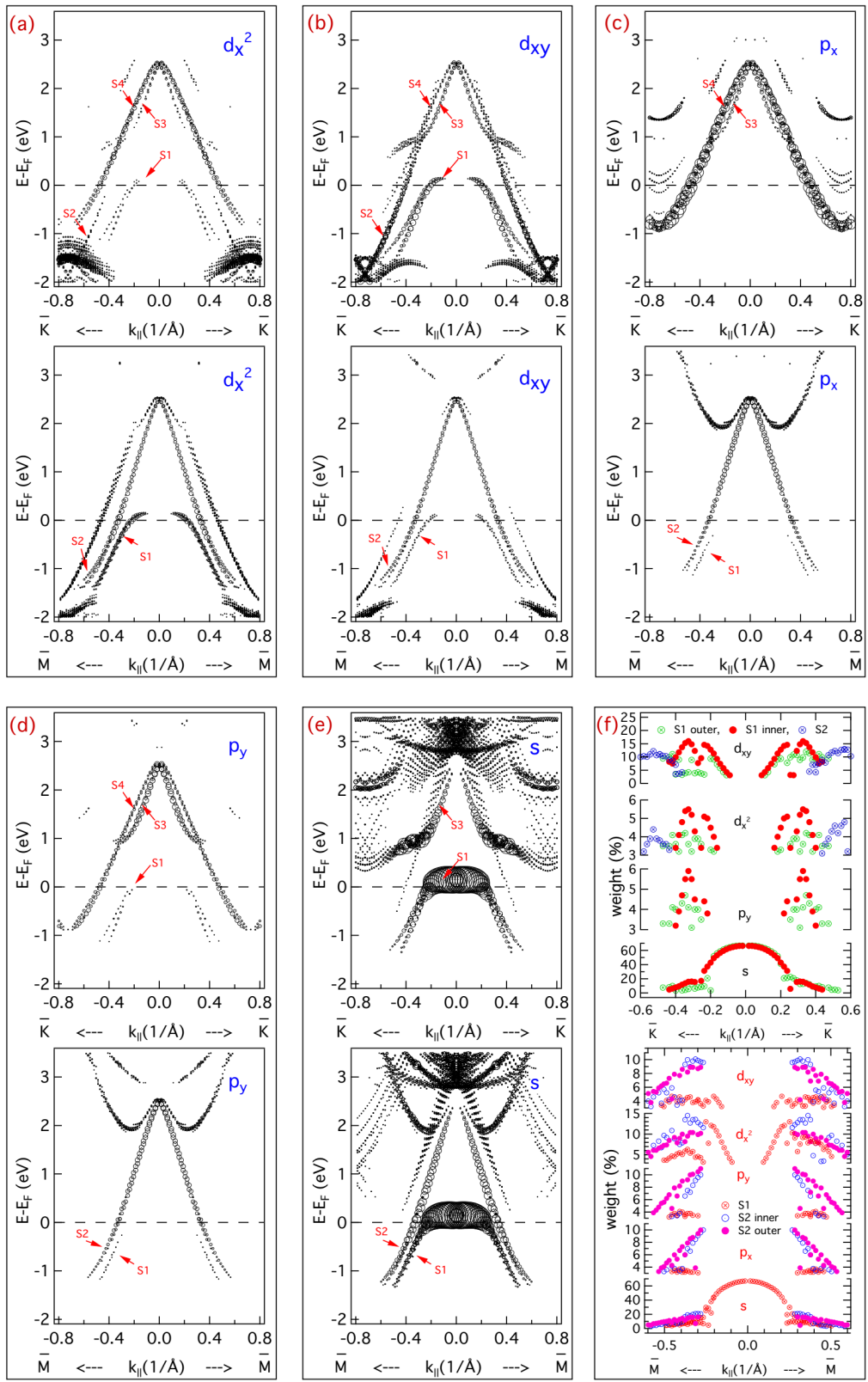

**Figure 5:** (color online)

The surface-, and orbital-projected band structure of the R3 phase of SnAu$_2$/Au(111). Panels (a), (b), (c), (d), and (e) show the total weight (proportional to size of the markers) of $d_x{}^2$, $d_{xy}$, $p_x$, $p_y$, and *s*-orbitals along the $\overline{\Gamma M'K}$ (top) and the $\overline{\Gamma K'M}$ (bottom) high symmetry directions, respectively. All these band structure projections contain only contributions originating from the topmost SnAu$_2$ alloy layer. (f) Spectral weights of the individual orbitals ($d_x{}^2$, $d_{xy}$, $p_x$, $p_y$, and *s*) contributing to the band S1 and S2 along $\overline{\Gamma M'K}$ (top) and $\overline{\Gamma K'M}$ (bottom) directions extracted from (a)–(e).

**Acknowledgement:**

The research leading to these results was financially supported by the German Science foundation (DFG) via SFB/TRR 173 Spin+X: spin in its collective environment (Project A02). M.M. acknowledges support by the Carl- Zeiss Stiftung for a post-doctoral fellowship. S.E. and B.S. thankfully acknowledge financial support from the Grad- uate School of Excellence MAINZ (Excellence Initiative DFG/GSC 266). S.M. acknowledges financial support via SFB1073 (Project B07). M.C. acknowledges funding from the European Research Council (ERC) under the European Union's Horizon 2020 research and innovation programme (grant agreement No. 725767 - hyControl).

# References

1. Molle, A., Goldberger, J., Houssa, M., Xu, Y., Zhang, S.-C. & Akinwande, D. Buckled two-dimensional Xene sheets. *Nat. Mater.* **16,** 163 (2017).

2. Balendhran, S., Walia, S., Nili, H., Sriram, S. & Bhaskaran, M. Elemental Analogues of Graphene: Silicene, Germanene, Stanene, and Phosphorene. *Small* **11,** 640–652 (2015).

3. Tersoff, J. Surface-Confined Alloy Formation in Immiscible Systems. *Phys. Rev. Lett.* **74,** 434–437 (1995).

4. Manchon, A., Koo, H. C., Nitta, J., Frolov, S. M. & Duine, R. A. New perspectives for Rashba spin-orbit coupling. *Nat Mater* **14,** 871–882 (2015).

5. Bhimanapati, G. R., Lin, Z., Meunier, V., Jung, Y., Cha, J., Das, S., Xiao, D., Son, Y., Strano, M. S., Cooper, V. R., Liang, L., Louie, S. G., Ringe, E., Zhou, W., Kim, S. S., Naik, R. R., Sumpter, B. G., Terrones, H., Xia, F., Wang, Y., Zhu, J., Akinwande, D., Alem, N., Schuller, J. A., Schaak, R. E., Terrones, M. & Robinson, J. A. Recent Advances in Two-Dimensional Materials beyond Graphene. *ACS Nano* **9,** 11509–11539 (2015).

6. Soumyanarayanan, A., Reyren, N., Fert, A. & Panagopoulos, C. Emergent phenomena induced by spin–orbit coupling at surfaces and interfaces. *Nature* **539,** 509–517 (2016).

7. Dil, J. H. Spin and angle resolved photoemission on non-magnetic low-dimensional systems. *J Phys Condens Matter* **21,** 403001 (2009).

8. Datta, S. & Das, B. Electronic analog of the electro-optic modulator. *Appl. Phys. Lett.* **56,** 665–667 (1990).

9. Sánchez, J. C. R., Vila, L., Desfonds, G., Gambarelli, S., Attané, J. P., Teresa, J. M. D., Magén, C. & Fert, A. Spin-to-charge conversion using Rashba coupling at the interface between non-magnetic materials. *Nat. Commun.* **4,** 2944 (2013).

10. Ast, C. R., Pacilé, D., Moreschini, L., Falub, M. C., Papagno, M., Kern, K., Grioni, M., Henk, J., Ernst, A., Ostanin, S. & Bruno, P. Spin-orbit split two-dimensional electron gas with tunable Rashba and Fermi energy. *Phys. Rev. B* **77,** 081407 (2008).

11. Meier, F., Petrov, V., Guerrero, S., Mudry, C., Patthey, L., Osterwalder, J. & Dil, J. H. Unconventional Fermi surface spin textures in the ${\text{Bi}}_{x}{\text{Pb}}_{1-x}/\text{Ag}(111)$ surface alloy. *Phys. Rev. B* **79,** 241408 (2009).

12. Gierz, I., Meier, F., Dil, J. H., Kern, K. & Ast, C. R. Tuning the spin texture in binary and ternary surface alloys on Ag(111). *Phys. Rev. B* **83,** 195122 (2011).

13. Ast, C. R., Henk, J., Ernst, A., Moreschini, L., Falub, M. C., Pacilé, D., Bruno, P., Kern, K. & Grioni, M. Giant Spin Splitting through Surface Alloying. *Phys. Rev. Lett.* **98,** 186807 (2007).

14. Pacilé, D., Ast, C. R., Papagno, M., Da Silva, C., Moreschini, L., Falub, M., Seitsonen, A. P. & Grioni, M. Electronic structure of an ordered Pb/Ag(111) surface alloy: Theory and experiment. *Phys. Rev. B* **73,** 245429 (2006).

15. Yaji, K., Ohtsubo, Y., Hatta, S., Okuyama, H., Miyamoto, K., Okuda, T., Kimura, A., Namatame, H., Taniguchi, M. & Aruga, T. Large Rashba spin splitting of a metallic surface-state band on a semiconductor surface. *Nat. Commun.* **1,** ncomms1016 (2010).

16. Varykhalov, A., Marchenko, D., Scholz, M. R., Rienks, E. D. L., Kim, T. K., Bihlmayer, G., Sánchez-Barriga, J. & Rader, O. Ir(111) Surface State with Giant Rashba Splitting Persists under Graphene in Air. *Phys. Rev. Lett.* **108,** 066804 (2012).

17. Stadtmüller, B., Seidel, J., Haag, N., Grad, L., Tusche, C., van Straaten, G., Franke, M., Kirschner, J., Kumpf, C., Cinchetti, M. & Aeschlimann, M. Modifying the Surface of a Rashba-Split Pb-Ag Alloy Using Tailored Metal-Organic Bonds. *Phys. Rev. Lett.* **117,** 096805 (2016).

18. Osiecki, J. R. & Uhrberg, R. I. G. Alloying of Sn in the surface layer of Ag(111). *Phys. Rev. B* **87,** 075441 (2013).

19. Bian, G., Wang, X., Miller, T. & Chiang, T.-C. Origin of giant Rashba spin splitting in Bi/Ag surface alloys. *Phys. Rev. B* **88,** 085427 (2013).

20. Moreschini, L., Bendounan, A., Gierz, I., Ast, C. R., Mirhosseini, H., Höchst, H., Kern, K., Henk, J., Ernst, A., Ostanin, S., Reinert, F. & Grioni, M. Assessing the atomic contribution to the Rashba spin-orbit splitting in surface alloys: Sb/Ag(111). *Phys. Rev. B* **79,** 075424 (2009).

21. Moreschini, L., Bendounan, A., Bentmann, H., Assig, M., Kern, K., Reinert, F., Henk, J., Ast, C. R. & Grioni, M. Influence of the substrate on the spin-orbit splitting in surface alloys on (111) noble-metal surfaces. *Phys. Rev. B* **80,** 035438 (2009).

22. Mirhosseini, H., Henk, J., Ernst, A., Ostanin, S., Chiang, C.-T., Yu, P., Winkelmann, A. & Kirschner, J. Unconventional spin topology in surface alloys with Rashba-type spin splitting. *Phys. Rev. B* **79,** 245428 (2009).

23. Meier, F., Petrov, V., Mirhosseini, H., Patthey, L., Henk, J., Osterwalder, J. & Dil, J. H. Interference of spin states in photoemission from Sb/Ag(111) surface alloys. *J. Phys. Condens. Matter* **23,** 072207 (2011).

24. Bentmann, H. & Reinert, F. Enhancing and reducing the Rashba-splitting at surfaces by adsorbates: Na and Xe on Bi/Cu(111). *New J. Phys.* **15,** 115011 (2013).

25. Bihlmayer, G., Blügel, S. & Chulkov, E. V. Enhanced Rashba spin-orbit splitting in Bi∕Ag(111) and Pb∕Ag(111) surface alloys from first principles. *Phys. Rev. B* **75,** 195414 (2007).

26. Jakobs, S., Ruffing, A., Jungkenn, D., Cinchetti, M., Mathias, S. & Aeschlimann, M. Spin structure of Rashba-split electronic states of Bi overlayers on Cu(111). *J Electron Spect Rel Phen* **201,** 47 (2015).

27. Mathias, S., Ruffing, A., Deicke, F., Wiesenmayer, M., Sakar, I., Bihlmayer, G., Chulkov, E. V., Koroteev, Y. M., Echenique, P. M., Bauer, M. & Aeschlimann, M. Quantum-Well-Induced Giant Spin-Orbit Splitting. *Phys. Rev. Lett.* **104,** 066802 (2010).

28. Bychkov, Y. A. & É. I. Rashba, P. *Zh Eksp Teor Fiz* **39,** 66 (1984).

29. Lee, H. & Choi, H. J. Role of $d$ orbitals in the Rashba-type spin splitting for noble-metal surfaces. *Phys. Rev. B* **86,** 045437 (2012).

30. Ishida, H. Rashba spin splitting of Shockley surface states on semi-infinite crystals. *Phys. Rev. B* **90,** 235422 (2014).

31. Noguchi, R., Kuroda, K., Yaji, K., Kobayashi, K., Sakano, M., Harasawa, A., Kondo, T., Komori, F. & Shin, S. Direct mapping of spin and orbital entangled wave functions under interband spin-orbit coupling of giant Rashba spin-split surface states. *Phys. Rev. B* **95,** 041111 (2017).

32. Simon, E., Szilva, A., Ujfalussy, B., Lazarovits, B., Zarand, G. & Szunyogh, L. Anisotropic Rashba splitting of surface states from the admixture of bulk states: Relativistic ab initio calculations and $k\ensuremath{\cdot}p$ perturbation theory. *Phys. Rev. B* **81,** 235438 (2010).

33. Bentmann, H., Forster, F., Bihlmayer, G., Chulkov, E. V., Moreschini, L., Grioni, M. & Reinert, F. Origin and manipulation of the Rashba splitting in surface alloys. *EPL Europhys. Lett.* **87,** 37003 (2009).

34. Gierz, I., Stadtmüller, B., Vuorinen, J., Lindroos, M., Meier, F., Dil, J. H., Kern, K. & Ast, C. R. Structural influence on the Rashba-type spin splitting in surface alloys. *Phys. Rev. B* **81,** 245430 (2010).

35. Corso, M., Verstraete, M. J., Schiller, F., Ormaza, M., Fernández, L., Greber, T., Torrent, M., Rubio, A. & Ortega, J. E. Rare-Earth Surface Alloying: A New Phase for GdAu_2. *Phys. Rev. Lett.* **105,** 016101 (2010).

36. Corso, M., Fernández, L., Schiller, F. & Ortega, J. E. Au(111)-Based Nanotemplates by Gd Alloying. *ACS Nano* **4,** 1603–1611 (2010).

37. Ormaza, M., Fernández, L., Ilyn, M., Magaña, A., Xu, B., Verstraete, M. J., Gastaldo, M., Valbuena, M. A., Gargiani, P., Mugarza, A., Ayuela, A., Vitali, L., Blanco-Rey, M., Schiller, F. & Ortega, J. E. High Temperature Ferromagnetism in a GdAg2 Monolayer. *Nano Lett.* **16,** 4230–4235 (2016).

38. Ormaza, M., Fernández, L., Lafuente, S., Corso, M., Schiller, F., Xu, B., Diakhate, M., Verstraete, M. J. & Ortega, J. E. LaAu_2 and CeAu_2 surface intermetallic compounds grown by high-temperature deposition on Au(111). *Phys. Rev. B* **88,** 125405 (2013).

39. Barth, J. V., Brune, H., Schuster, R., Ertl, G. & Behm, R. J. Intermixing and two-dimensional alloy formation in the Na/Au(111) system. *Surf. Sci.* **292,** L769–L774 (1993).

40. Chen, W.-C., Chang, T.-R., Tsai, S.-T., Yamamoto, S., Kuo, J.-M., Cheng, C.-M., Ku-Ding Tsuei, Yaji, K., Lin, H., Jeng, H.-T., Mou, C.-Y., Matsuda, I. & Tang, S.-J. Significantly enhanced giant Rashba splitting in a thin film of binary alloy. *New J. Phys.* **17,** 083015 (2015).

41. Crepaldi, A., Pons, S., Frantzeskakis, E., Calleja, F., Etzkorn, M., Seitsonen, A. P., Kern, K., Brune, H. & Grioni, M. Combined ARPES and STM study of Pb/Au(111) Moir\'e structure: One overlayer, two symmetries. *Phys. Rev. B* **87,** 115138 (2013).

42. Bihlmayer, G., Koroteev, Y. M., Echenique, P. M., Chulkov, E. V. & Blügel, S. The Rashba-effect at metallic surfaces. *Surf. Sci.* **600,** 3888–3891 (2006).

43. Mathias, S., Ruffing, A., Deicke, F., Wiesenmayer, M., Aeschlimann, M. & Bauer, M. Band structure dependence of hot-electron lifetimes in a Pb/Cu(111) quantum-well system. *Phys. Rev. B* **81,** 155429 (2010).

44. LaShell, S., McDougall, B. A. & Jensen, E. Spin Splitting of an Au(111) Surface State Band Observed with Angle Resolved Photoelectron Spectroscopy. *Phys Rev Lett* **77,** 3419 (1996).

45. Shukla, A. K., Banik, S., Dhaka, R. S., Biswas, C., Barman, S. R. & Haak, H. Versatile UHV compatible Knudsen type effusion cell. *Rev Sci Instrum* **75,** 4467 (2004).

46. Maniraj, M., Rai, A., Barman, S. R., Krajčí, M., Schlagel, D. L., Lograsso, T. A. & Horn, K. Unoccupied electronic states of icosahedral Al-Pd-Mn quasicrystals: Evidence of image potential resonance and pseudogap. *Phys. Rev. B* **90,** 115407 (2014).

47. Kollamana, J., Wei, Z., Lyu, L., Zimmer, M., Dietrich, F., Eul, T., Stöckl, J., Maniraj, M., Ponzoni, S., Cinchetti, M., Stadtmüller, B., Gerhards, M. & Aeschlimann, M. Control of Cooperativity through a Reversible Structural Phase Transition in MoMo-Methyl/Cu(111). *Adv. Funct. Mater.* n/a-n/a doi:10.1002/adfm.201703544

48. Kollamana, J., Wei, Z., Laux, M., Stöckl, J., Stadtmüller, B., Cinchetti, M. & Aeschlimann, M. Scanning Tunneling Microscopy Study of Ordered C60 Submonolayer Films on Co/Au(111). *J. Phys. Chem. C* **120,** 7568–7574 (2016).

49. Horcas, I., Fernández, R., Gómez-Rodríguez, J. M., Colchero, J., Gómez-Herrero, J. & Baro, A. M. WSXM: A software for scanning probe microscopy and a tool for nanotechnology. *Rev. Sci. Instrum.* **78,** 013705 (2007).

50. Kresse, G. & Furthmüller, J. Efficient iterative schemes for ab initio total-energy calculations using a plane-wave basis set. *Phys. Rev. B* **54,** 11169–11186 (1996).

51. Blöchl, P. E. Projector augmented-wave method. *Phys. Rev. B* **50,** 17953–17979 (1994).

52. Kresse, G. & Joubert, D. From ultrasoft pseudopotentials to the projector augmented-wave method. *Phys. Rev. B* **59,** 1758–1775 (1999).

53. Perdew, J. P., Burke, K. & Ernzerhof, M. Generalized Gradient Approximation Made Simple. *Phys. Rev. Lett.* **77,** 3865–3868 (1996).

54. Harten, U., Lahee, A. M., Toennies, J. P. & Wöll, C. Observation of a Soliton Reconstruction of Au(111) by High-Resolution Helium-Atom Diffraction. *Phys. Rev. Lett.* **54,** 2619–2622 (1985).

55. Wöll, C., Chiang, S., Wilson, R. J. & Lippel, P. H. Determination of atom positions at stacking-fault dislocations on Au(111) by scanning tunneling microscopy. *Phys. Rev. B* **39,** 7988–7991 (1989).

56. Barth, J. V., Brune, H., Ertl, G. & Behm, R. J. Scanning tunneling microscopy observations on the reconstructed Au(111) surface: Atomic structure, long-range superstructure, rotational domains, and surface defects. *Phys. Rev. B* **42,** 9307–9318 (1990).

57. Bürgi, L., Brune, H. & Kern, K. Imaging of Electron Potential Landscapes on Au(111). *Phys. Rev. Lett.* **89,** 176801 (2002).

58. Narasimhan, S. & Vanderbilt, D. Elastic stress domains and the herringbone reconstruction on Au(111). *Phys. Rev. Lett.* **69,** 1564–1567 (1992).

59. Hanke, F. & Björk, J. Structure and local reactivity of the Au(111) surface reconstruction. *Phys. Rev. B* **87,** 235422 (2013).

60. Lauwaet, K., Schouteden, K., Janssens, E., Van Haesendonck, C., Lievens, P., Trioni, M. I., Giordano, L. & Pacchioni, G. Resolving all atoms of an alkali halide via nanomodulation of the thin NaCl film surface using the Au(111) reconstruction. *Phys. Rev. B* **85,** 245440 (2012).

61. Kawai, S., Benassi, A., Gnecco, E., Söde, H., Pawlak, R., Feng, X., Müllen, K., Passerone, D., Pignedoli, C. A., Ruffieux, P., Fasel, R. & Meyer, E. Superlubricity of graphene nanoribbons on gold surfaces. *Science* **351,** 957–961 (2016).

62. Roussel, T. J., Barrena, E., Ocal, C. & Faraudo, J. Predicting supramolecular self-assembly on reconstructed metal surfaces. *Nanoscale* **6,** 7991–8001 (2014).

63. Chambliss, D. D., Wilson, R. J. & Chiang, S. Nucleation of ordered Ni island arrays on Au(111) by surface-lattice dislocations. *Phys. Rev. Lett.* **66,** 1721–1724 (1991).

64. Barth, J. V., Costantini, G. & Kern, K. Engineering atomic and molecular nanostructures at surfaces. *Nature* (2005). doi:10.1038/nature04166

65. Bulou, H. & Bucher, J.-P. Long Range Substrate Mediated Mass Transport on Metal Surfaces Induced by Adatom Clusters. *Phys. Rev. Lett.* **96,** 076102 (2006).

66. Casari, C. S., Foglio, S., Siviero, F., Li Bassi, A., Passoni, M. & Bottani, C. E. Direct observation of the basic mechanisms of Pd island nucleation on Au(111). *Phys. Rev. B* **79,** 195402 (2009).

67. Nenchev, G., Diaconescu, B., Hagelberg, F. & Pohl, K. Self-assembly of methanethiol on the reconstructed Au(111) surface. *Phys. Rev. B* **80,** 081401 (2009).

68. Bian, G., Zhang, L., Liu, Y., Miller, T. & Chiang, T.-C. Illuminating the Surface Spin Texture of the Giant-Rashba Quantum-Well System $\mathrm{Bi}/\mathrm{Ag}(111)$ by Circularly Polarized Photoemission. *Phys. Rev. Lett.* **108,** 186403 (2012).

69. Reinert, F. Spin–orbit interaction in the photoemission spectra of noble metal surface states. *J. Phys. Condens. Matter* **15,** S693 (2003).

70. Cercellier, H., Fagot-Revurat, Y., Kierren, B., Reinert, F., Popović, D. & Malterre, D. Spin-orbit splitting of the Shockley state in the Ag∕Au(111) interface. *Phys. Rev. B* **70,** 193412 (2004).

71. Winkelmann, A., Ünal, A. A., Tusche, C., Ellguth, M., Chiang, C.-T. & Kirschner, J. Direct k -space imaging of Mahan cones at clean and Bi-covered Cu(111) surfaces. *New J. Phys.* **14,** 083027 (2012).

72. Maniraj, M., D'Souza, S. W., Rai, A., Schlagel, D. L., Lograsso, T. A., Chakrabarti, A. & Barman, S. R. Unoccupied electronic structure of Ni2MnGa ferromagnetic shape memory alloy. *Solid State Commun.* **222,** 1–4 (2015).

73. Maniraj, M., D'Souza, S. W., Singh, S., Biswas, C., Majumdar, S. & Barman, S. R. Inverse photoemission and photoemission spectroscopic studies on sputter-annealed Ni–Mn–Sn and Ni–Mn–In surfaces. *J. Electron Spectrosc. Relat. Phenom.* **197,** 106–111 (2014).

74. Bentmann, H., Kuzumaki, T., Bihlmayer, G., Blügel, S., Chulkov, E. V., Reinert, F. & Sakamoto, K. Spin orientation and sign of the Rashba splitting in Bi/Cu(111). *Phys. Rev. B* **84,** 115426 (2011).

75. Meier, F., Dil, H., Lobo-Checa, J., Patthey, L. & Osterwalder, J. Quantitative vectorial spin analysis in angle-resolved photoemission: $\mathrm{Bi}/\mathrm{Ag}(111)$ and $\mathrm{Pb}/\mathrm{Ag}(111)$. *Phys. Rev. B* **77,** 165431 (2008).

76. Bondarenko, L. V., Gruznev, D. V., Yakovlev, A. A., Tupchaya, A. Y., Usachov, D., Vilkov, O., Fedorov, A., Vyalikh, D. V., Eremeev, S. V., Chulkov, E. V., Zotov, A. V. & Saranin, A. A. Large spin splitting of metallic surface-state bands at adsorbate-modified gold/silicon surfaces. *Sci. Rep.* **3,** 1826 (2013).

77. Frantzeskakis, E. & Grioni, M. Anisotropy effects on Rashba and topological insulator spin-polarized surface states: A unified phenomenological description. *Phys. Rev. B* **84,** 155453 (2011).

78. Gruznev, D. V., Bondarenko, L. V., Matetskiy, A. V., Yakovlev, A. A., Tupchaya, A. Y., Eremeev, S. V., Chulkov, E. V., Chou, J.-P., Wei, C.-M., Lai, M.-Y., Wang, Y.-L., Zotov, A. V. & Saranin, A. A. A Strategy to Create Spin-Split Metallic Bands on Silicon Using a Dense Alloy Layer. *Sci. Rep.* **4,** 4742 (2014).

79. Premper, J., Trautmann, M., Henk, J. & Bruno, P. Spin-orbit splitting in an anisotropic two-dimensional electron gas. *Phys. Rev. B* **76,** 073310 (2007).

80. Fu, L. Hexagonal Warping Effects in the Surface States of the Topological Insulator ${\mathrm{Bi}}_{2}{\mathrm{Te}}_{3}$. *Phys. Rev. Lett.* **103,** 266801 (2009).

81. Höpfner, P., Schäfer, J., Fleszar, A., Dil, J. H., Slomski, B., Meier, F., Loho, C., Blumenstein, C., Patthey, L., Hanke, W. & Claessen, R. Three-Dimensional Spin Rotations at the Fermi Surface of a Strongly Spin-Orbit Coupled Surface System. *Phys. Rev. Lett.* **108,** 186801 (2012).

82. Gong, S.-J., Cai, J., Yao, Q.-F., Tong, W.-Y., Wan, X., Duan, C.-G. & Chu, J. H. Orbital control of Rashba spin orbit coupling in noble metal surfaces. *J. Appl. Phys.* **119,** 125310 (2016).

83. Reinert, F. & Nicolay, G. Influence of the herringbone reconstruction on the surface electronic structure of Au(111). *Appl. Phys. A* **78,** 817–821 (2004).

84. Hochstrasser, M., Tobin, J. G., Rotenberg, E. & Kevan, S. D. Spin-Resolved Photoemission of Surface States of W(110)−(1x1)H. *Phys. Rev. Lett.* **89,** 216802 (2002).

85. Hortamani, M. & Wiesendanger, R. Role of hybridization in the Rashba splitting of noble metal monolayers on W(110). *Phys. Rev. B* **86,** 235437 (2012).

86. Rotenberg, E., Chung, J. W. & Kevan, S. D. Spin-Orbit Coupling Induced Surface Band Splitting in Li/W(110) and Li/Mo(110). *Phys. Rev. Lett.* **82,** 4066–4069 (1999).

87. Rybkin, A. G., Shikin, A. M., Adamchuk, V. K., Marchenko, D., Biswas, C., Varykhalov, A. & Rader, O. Large spin-orbit splitting in light quantum films: Al/W(110). *Phys. Rev. B* **82,** 233403 (2010).

88. Miyamoto, K., Kimura, A., Kuroda, K., Okuda, T., Shimada, K., Namatame, H., Taniguchi, M. & Donath, M. Spin-Polarized Dirac-Cone-Like Surface State with $d$ Character at W(110). *Phys. Rev. Lett.* **108,** 066808 (2012).